\documentstyle[preprint,eqsecnum,aps]{revtex}

\begin{document}
\draft
%\preprint{}
\title{Magnetization process of the spin-1/2 $XXZ$ models\\
on square and cubic lattices
}
\author{Masanori Kohno and Minoru Takahashi}
\address{
Institute for Solid State Physics,\\
University of Tokyo, Roppongi, Minato-ku, Tokyo 106
}
\date{\today}
\maketitle
\begin{abstract}
The magnetization process of the spin-1/2 antiferromagnetic $XXZ$ model 
with Ising-like anisotropy in the ground state is investigated. 
We show numerically that the Ising-like $XXZ$ models 
on square and cubic lattices show a first-order phase transition 
at some critical magnetic field. We estimate the value of the critical field 
and the magnetization jump on the basis of the Maxwell construction. 
The magnetization jump in the Ising-limit is investigated 
by means of perturbation theory. 
Based on our numerical results, we briefly discuss the phase diagram 
of the extended Bose-Hubbard model in the hard-core limit.
\end{abstract}
\pacs{PACS numbers: 75.30.Kz, 75.40.Cx, 05.30.Jp, 67.40.Db}

\narrowtext

\section{Introduction}
\label{Introduction}

N\'eel\cite{Neel} predicted the first-order transition of anisotropic antiferromagnets in the presence of a magnetic field in 1936. 
He pointed out that the spins will abruptly change directions 
from parallel to perpendicular with respect to the $c$-axis 
(easy axis of sublattice magnetization) 
at some value of the external magnetic field, when the magnetic field is applied 
to the direction parallel to the $c$-axis.
His prediction was confirmed experimentally\cite{experiment}, 
and this first-order phase transition is now known as the spin-flopping process.
Thirty years later than the discovery of the spin-flopping process,
C.N. Yang and C.P. Yang showed by the Bethe ansatz that the one-dimensional (1d) 
spin-1/2 $XXZ$ model with Ising-like anisotropy exhibits a second-order transition 
in the presence of a magnetic field\cite{YY}.  
Thus, for one-dimensional Ising-like antiferromagnets, quantum fluctuations, 
which are neglected in the mean-field approximation, play an essential role.
In this way, quantum fluctuations may drastically modify the classical behavior 
depending on the dimensionality.
Hence, we investigate the magnetization process 
of the spin-1/2 Ising-like $XXZ$ (I-$XXZ$) models in two and three dimensions  
in order to see how quantum fluctuations modify the classical behavior of 
the magnetization process of Ising-like antiferromagnets.
\par
Also, the spin-1/2 $XXZ$ model can be translated into the hard-core boson model 
with nearest neighbor repulsion\cite{matsu2}.
This model corresponds to a special case of the extended Bose-Hubbard model 
which is considered to be relevant for low-temperature properties of 
liquid helium on a periodic substrate and also for Josephson junction arrays
\cite{bose_Hub_scal1,bose_Hub_scal2,bose_Hub_mf1}.
From the theoretical point of view, a lot of attention has been paid
to the Bose-Hubbard model as the simplest model to describe 
the superfluid-insulator transition
\cite{bose_Hub_scal1,bose_Hub_scal2,bose_Hub_mf1,bose_Hub_QMC2,bose_Hub_QMC3}.
We can obtain information about the superfluid-insulator transition 
occurring in the extended Bose-Hubbard model 
through the investigation of the spin-1/2 I-$XXZ$ model.
\par
This paper is organized as follows: 
In Sec. \ref{Model}, the $XXZ$ model is defined.
The details of the numerical calculations are presented.
In Sec. \ref{Classical}, we review the classical Ising-like $XXZ$ model.
Numerical results on the magnetization curve of the spin-1/2 I-$XXZ$ models 
in two and three dimensions are shown in Sec. \ref{Spin-1/2}.
In Sec. \ref{Ising-limit}, the behavior of the magnetization curve 
in the Ising-limit is investigated by means of perturbation theory.
In Sec. \ref{Bose-Hubbard}, we briefly discuss 
the superfluid-insulator transition 
in the extended Bose-Hubbard model in the hard-core limit 
based on the numerical results of the spin-1/2 I-$XXZ$ model.
Section \ref{Summary} is devoted to summary.

\section{Model and Method}
\label{Model}

In the present paper, we consider the $XXZ$ model defined by the following Hamiltonian:
\begin{equation}
   {\cal H}_{XXZ} = J\sum_{\langle i,j\rangle}
                    (S^{x}_{i}S^{x}_{j}+S^{y}_{i}S^{y}_{j} 
                     + \lambda S^{z}_{i}S^{z}_{j}),
\end{equation}
where $S^{x(y,z)}_{i}$ denote the $x(y,z)$ components of the spin operator 
at site $i$. Here $\langle i,j\rangle$ denotes nearest neighbors. 
The anisotropic coupling constant is denoted by $\lambda$. 
For $\lambda=1$, the isotropic Heisenberg model is recovered.
We investigate the spin-1/2 $XXZ$ models on square and cubic lattices 
in the ground state in the canonical ensemble.
Namely, we measure the energy $E$ within the subspace 
of fixed magnetization $M$ (=$\sum_i S^{z}_i$).
The magnetic field in a finite-size cluster is defined 
as $H(\bar M)\equiv (E(M_1)-E(M_2))/(M_1-M_2)$, 
where $\bar M$ is defined by $(M_1+M_2)/2$. 
In the thermodynamic limit, this definition of the magnetic field reduces 
to the normal one: $H\equiv \partial{E}/\partial{M}$.
The maximum magnetization and the saturation field are denoted 
by $M_{\rm max}$(=$N_{\rm s}/2$) and $H_{\rm max}$(=$J(\lambda+1)d$), respectively.
Here $N_{\rm s}$ and $d$ are the system size and the spatial dimensionality.
\par
We use the Lanczos algorithm (exact diagonalization) 
for clusters up to 32 sites 
and the cluster algorithm\cite{clst_alg} (quantum Monte Carlo) 
for larger clusters up to 100 sites. 
For the cluster algorithm, the measurements have been performed 
at the inverse temperature $\beta J=16$. 
The width of the Trotter slice is chosen as $\Delta\tau J=0.04$ for two dimensions (2d) 
and $\Delta\tau J=0.053$ for three dimensions (3d). 
The simulation has been performed in the canonical ensemble. 
In the small $\lambda$ regime, the number of points near $M=0$ 
obtained by exact diagonalization is too small to use the Maxwell construction.
On the other hand, in the large $\lambda$ regime, 
statistical errors in the quantum Monte Carlo calculation become large, 
because the spin configurations are almost frozen.
Hence, we have investigated the small $\lambda$ regime by quantum Monte Carlo 
and the large $\lambda$ regime by exact diagonalization.
\par
In order to see finite-size effects, we show the size-dependence of 
the energy gap $\Delta_{\rm g}(\equiv E(M=1)-E(M=0))$, 
the critical field $H_{\rm c}$ and the magnetization jump $M_{\rm s}$ 
in Fig.\ref{size}.
As shown in this figure, the size-dependence is very small. 
As a check of numerical accuracy, we compare the energy gap $\Delta_{\rm g}$ 
obtained by this method with those obtained by other methods 
as shown in Fig.\ref{gap}.
Our result is quite consistent with those 
of third-order spin-wave theory\cite{3OSW} 
and the series expansion around the Ising-limit\cite{expansion}.
Hence, we consider that the inverse temperature $\beta$ 
and the width of the Trotter slice $\Delta\tau$ are sufficient. 
[See also Fig.\ref{mh}.]
For three dimensions, we report the 64-site results.

\section{Review of the Classical Spin Case}
\label{Classical}

Before investigating the spin-1/2 $XXZ$ models, we briefly review 
the magnetization process of the classical I-$XXZ$ model in the ground state\cite{Neel,Yosida}. 
The ground state energy of the classical I-$XXZ$ model 
may be written in the following form:
\begin{eqnarray}
   E_{{\rm C}-XXZ} &=& \frac{JN_{\rm s}z}{2}
                          (S^{x}_{A}S^{x}_{B}+S^{y}_{A}S^{y}_{B} 
                           + \lambda S^{z}_{A}S^{z}_{B})
\nonumber\\
                   &=& -\frac{JN_{\rm s}zS^2}{2}
                          (\sin(\theta+\phi)\sin(\theta-\phi) 
\nonumber\\
                    &&+ \lambda \cos(\theta+\phi)\cos(\theta-\phi)),
\end{eqnarray}
where $S^{x(y,z)}_{A(B)}$ represent the $x(y,z)$ components 
of the spin operators at a site in the A(B) sublattices. 
The length of the spin and the coordination number are denoted by $S$ and $z(=2d)$, 
respectively. 
The angles $\theta$ and $\phi$ are defined as in Fig.\ref{class}(a) 
($0\le\theta\le\pi/2$, $0\le\phi\le\pi/2$).
The Zeeman term $E_{\rm Z}$ is written in the following form:
\begin{equation}
   E_{\rm Z} = -\frac{HN_{\rm s}}{2}(S^{z}_{A}+S^{z}_{B})
             = \frac{HN_{\rm s}S}{2}(\cos(\theta+\phi)-\cos(\theta-\phi)).
\end{equation}
By minimizing the total energy ($E_{\rm tot}\equiv E_{{\rm C}-XXZ}+E_{\rm Z}$) 
with respect to $\theta$ and $\phi$, one finds the following stable states (Fig.\ref{class}(b)):\\
\begin{tabular}{clllcl}
(i) &$\theta=0$,&$\phi=0$,& $E_{\rm tot}=-{\tilde J}\lambda$ &for& 
${\tilde H}<{\tilde J}\sqrt{\lambda^2-1}$\\
(ii) &$\theta=\pi/2$,&$\phi=\arcsin\frac{{\tilde H}}{{\tilde J}(\lambda+1)}$,& 
$E_{\rm tot}=-{\tilde J}-\frac{{\tilde H}^2}{{\tilde J}(\lambda+1)}$ 
&for& ${\tilde J}(\lambda+1)>{\tilde H}>{\tilde J}\sqrt{\lambda^2-1}$\\
(iii) &$\theta=\pi/2$,&$\phi=\pi/2$,& 
$E_{\rm tot}={\tilde J}\lambda-2{\tilde H}$ 
&for& ${\tilde H}>{\tilde J}(\lambda+1)$,
\end{tabular}\\
where ${\tilde J}$ and ${\tilde H}$ are defined 
as ${\tilde J}\equiv JN_{\rm s}zS^2/2$ and ${\tilde H}\equiv HN_{\rm s}S/2$.
The magnetization curve of the classical I-$XXZ$ model is shown in Fig.\ref{class}(c).
The transition from the state (i) to the state (ii) is known as 
the spin-flopping process\cite{Neel,Yosida}.
The critical field $H_{\rm c}$ is defined as the magnetic field 
above which the ground state has non-zero magnetization.  
The $\lambda$ dependence of the critical field $H_{\rm c}$ 
and that of the magnetization jump $M_{\rm s}$ are obtained as
\begin{equation}
   H_{\rm c}/H_{\rm max} = 
   M_{\rm s}/M_{\rm max} =  \sqrt{(\lambda-1)/(\lambda+1)},
\label{Hc_classical}
\end{equation}
where $H_{\rm max}=JSz(\lambda+1)$ and $M_{\rm max}=N_{\rm s}S$.

\section{Magnetization Curve of the Spin-1/2 $XXZ$ Model}
\label{Spin-1/2}

In this section, we present the numerical results 
on the magnetization curve of the spin-1/2 I-$XXZ$ models on square and cubic lattices.
As an example, we show the magnetization curve of the spin-1/2 $XXZ$ model 
for $\lambda=2$ on a square lattice in Fig. \ref{mh}.
The critical field $H_{\rm c}$ and the magnetization jump $M_{\rm s}$ 
are determined on the basis of the Maxwell construction as follows. 
[See also Fig.\ref{Ising}(a).]
We fit the energy as a function of magnetization by a polynomial. 
The tangent from the point at $M=0$ to the fitting curve gives 
a lower energy than the numerical data in the region of $0<M<M_{\rm s}$. 
Here $M_{\rm s}$ is the magnetization at the point of contact 
between the fitting curve and the tangent. 
Hence, we can identify the region of phase separation 
as $0<M<M_{\rm s}$ on the basis of the Maxwell construction
\cite{ene_size_correction}. 
The magnetic field of the phase-separated state ($H_{\rm c}$) is given 
as the slope of the tangent.
In practice, we have determined the phase-separation boundary as the point 
where the following condition is satisfied:
$\partial{E}/\partial{M}=(E(M)-E(M=0))/M$ as in Fig.\ref{det}.
The size-dependence of the critical field $H_{\rm c}$ 
and that of the magnetization jump $M_{\rm s}$ determined 
by the Maxwell construction are very small 
as discussed in Sec.\ref{Model} (Fig.\ref{size}).
\par
We show the $\lambda$ dependence of the critical field $H_{\rm c}$ 
in Fig.\ref{hc}.
The critical field $H_{\rm c}$ is suppressed by quantum fluctuations.
In order to see how large the critical field $H_{\rm c}$ is suppressed 
by quantum fluctuations, we have tried to fit the numerical data as 
$H_{\rm c}/H_{\rm max}=(\frac{\lambda-1}{\lambda+1})^{\alpha}$, 
analogously with the classical result $\alpha=0.5$ (eq.(\ref{Hc_classical})).
We estimate $\alpha$ to be $\alpha=0.64\pm0.01$ for 2d 
and $\alpha=0.57\pm0.01$ for 3d.
Note that the $\lambda$ dependence of the critical field $H_{\rm c}$ 
in two and three dimensions is quite different from the one-dimensional case, 
where the gap
(=$H_{\rm c}\equiv\partial{E}/\partial{M}|_{M/M_{\rm max}\rightarrow+0}$) 
opens exponentially: 
$H_{\rm c}\propto\exp[-\pi^2/2\sqrt{2(\lambda-1)}]$\cite{dG}.
\par
Here, we mention the relation between the critical field $H_{\rm c}$ 
and the energy gap $\Delta_{\rm g}$.
It is expected that the energy gap $\Delta_{\rm g}$ is larger 
than the critical field $H_{\rm c}$, if a first-order transition occurs 
in the presence of a magnetic field. The reason is as follows.
The ground state of $M=1$ is considered to be the one-magnon state, 
which may be described by spin-wave theory.
Hence, the gap $\Delta_{\rm g}$ corresponds to the excitation energy of one magnon
from the ground state of $M=0$.
On the other hand, phase separation occurs, because magnons gain energy by interacting attractively with each other.
The critical field $H_{\rm c}$ would be determined 
by the effective attractive interactions 
between the macroscopic number of magnons.
As a result, if phase separation occurs, 
the gap $\Delta_{\rm g}$ is expected to be larger than the critical field $H_{\rm c}$, i.e.
\begin{equation}
\Delta_{\rm g}>H_{\rm c}\equiv\partial{E}/\partial{M}|_{M/M_{\rm max}\rightarrow+0},
\end{equation}
where $M$ is assumed to be a macroscopic number 
when the limit $M/M_{\rm max}\rightarrow+0$ is taken.
We compare the gap $\Delta_{\rm g}$ and the critical field $H_{\rm c}$ 
of the spin-1/2 I-$XXZ$ model on a square lattice in Fig.\ref{gap_Hc}.
The gap $\Delta_{\rm g}$ is always larger than the critical field $H_{\rm c}$ as expected.
It is interesting to contrast this behavior with the one-dimensional result.
For one dimension, the transition is of second order\cite{YY},
and the following relation is satisfied:
$\partial{E}/\partial{M}|_{M/M_{\rm max}\rightarrow+0}=E(M=1)-E(M=0)$.
This is considered to be due to effective repulsive interactions.
\par
The $\lambda$ dependence of the magnetization jump $M_{\rm s}$ 
is shown in Fig.\ref{ms}.
We estimate the critical value of $\lambda$, where $M_{\rm s}$ vanishes, as 
$\lambda_{\rm c}=1.00\pm 0.02$ by extrapolating the data in Fig.\ref{ms}.
This confirms that the spin-1/2 I-$XXZ$ models
on square and cubic lattices show a first-order transition 
at some critical field for any value 
of the anisotropic coupling constant larger than one ($\lambda>1$).
The $\lambda$ dependence of the magnetization jump $M_{\rm s}$ is remarkably different from the classical result, especially in the large $\lambda$ regime.

\section{Ising-limit}
\label{Ising-limit}

In this section, we discuss the magnetization process 
of the spin-1/2 $XXZ$ model in the Ising-limit. 
In Fig.\ref{ms}, the value of the magnetization jump $M_{\rm s}$ 
in the Ising-limit 
($\lambda\rightarrow\infty$) does not coincide 
with that of the Ising model ($\lambda=\infty$)($M_{\rm s}(\lambda=\infty)=M_{\rm max}$): 
\begin{equation}
M_{\rm s}(\lambda\rightarrow\infty)\ne M_{\rm s}(\lambda=\infty).
\end{equation}
This can be explained by means of perturbation theory as follows. 
We rewrite the spin-1/2 $XXZ$ model as 
\begin{equation}
   {\cal H}_{XXZ} = {\bar J}\sum_{\langle i,j\rangle}S^{z}_{i}S^{z}_{j}
 + \epsilon{\bar J}\sum_{\langle i,j\rangle}
(S^{x}_{i}S^{x}_{j}+S^{y}_{i}S^{y}_{j}), 
\end{equation}
where ${\bar J}$ and $\epsilon$ are defined as ${\bar J}\equiv J\lambda$ 
and $\epsilon\equiv 1/\lambda$. 
We consider the $XY$-term as the perturbation.
At $M=0$, the unperturbed ground states are the two-degenerate N\'eel states. 
The leading perturbation energy is of order $\epsilon^2$. 
On the other hand, in the limit of $M\rightarrow M_{\rm max}$, 
the leading perturbation energy is of order $\epsilon$ 
and proportional to $M-M_{\rm max}$. 
Hence it is expected that phase separation occurs 
for the magnetization smaller than some value $M_{\rm s}$($<M_{\rm max}$) 
in the Ising-limit.
We numerically estimate the value of $M_{\rm s}$ in the Ising-limit 
with first-order perturbation theory in the following way.
The first-order perturbation energy $E_1$ is obtained as  
\begin{equation}
   E_1 = \frac{\epsilon{\bar J}}{2}\frac{\sum_{\alpha,\beta}\langle \beta|
\sum_{\langle i,j\rangle}(S^{+}_{i}S^{-}_{j}+S^{+}_{i}S^{-}_{j})|\alpha \rangle}
{\sum_{\alpha}\langle \alpha|\alpha \rangle}, 
\end{equation}
where $|\alpha\rangle$ and $|\beta\rangle$ denote unperturbed ground states 
of the Ising model in the subspace of fixed magnetization. 
We generate $|\alpha\rangle$'s randomly and measure $E_1$ 
using Monte Carlo technique. The value of $M_{\rm s}$ is determined 
based on the Maxwell construction.
Figure \ref{Ising} shows the first-order perturbation energy $E_1$ 
and the value of $M_{\rm s}$ in the Ising-limit 
on hyper-cubic lattices in dimensions up to six.
We extrapolate the data in Fig.\ref{Ising}(b) and estimate the inverse dimensionality, 
where $M_{\rm s}$ coincides with $M_{\rm max}$, as $1/d=0.01\pm 0.02$. 
This confirms that the value of $M_{\rm s}$ in the Ising-limit
($\lambda\rightarrow\infty$) does not coincide 
with that of the Ising model ($\lambda=\infty$) in finite spatial dimensions.

\section{Relation to the Bose-Hubbard Model}
\label{Bose-Hubbard}

In this section, we briefly discuss the phase diagram 
of the hard-core boson model with nearest-neighbor repulsion 
based on the numerical results of the spin-1/2 $XXZ$ models.
The hard-core boson model with nearest-neighbor repulsion can be obtained 
from the following extended Bose-Hubbard model 
by taking the on-site repulsion $U$ infinity:
\begin{equation}
   {\cal H}_{\rm BH} = t \sum_{\langle i,j\rangle}(b^{\dag}_{i}b_{j}
                                     +b^{\dag}_{j}b_{i}) 
                      + U \sum_{i}n_{i}(n_{i}-1)
                      + V \sum_{\langle i,j\rangle}(n_{i}-1/2)(n_{j}-1/2),
\end{equation}
where $b^{\dag}_{i}$ ($b_{i}$) creates (annihilates) a boson 
on site $i$ and $n_{i}=b^{\dag}_{i}b_{i}$. 
Here $\langle i,j\rangle$ denotes nearest neighbors.
The spin-1/2 $XXZ$ model can be translated 
into the hard-core boson model with nearest-neighbor repulsion
($t\leftrightarrow J/2$, $V\leftrightarrow J\lambda/4$)\cite{matsu2}.
Figure \ref{ms} corresponds to the phase diagram of this model 
by relating the filling $\rho$ and the interaction strength $V/t$ 
to $M/M_{\rm max}$ and $\lambda$ 
according to $\rho=(1-M/M_{\rm max})/2$ and $V/t=\lambda/2$. 
The numerical results in the previous section are translated as follows:
The superfluid-insulator transition occurs in the region of $V>t/2$.
This transition is a first-order transition, 
which is consistent with recent investigation 
of the Bose-Hubbard model\cite{bose_Hub_QMC3}.
Phase separation does not occur for the density $\rho$ smaller 
than $\rho_{\rm c}$, for finite $V/t$, 
where $\rho_{\rm c}\equiv 
(1-M_{\rm s}(\lambda\rightarrow\infty)/M_{\rm max})/2>0$.
This $\rho_{\rm c}$ approaches zero as the spatial dimensionality $d$ 
goes to infinity (Fig.\ref{Ising}(b)).

\section{Summary}
\label{Summary}

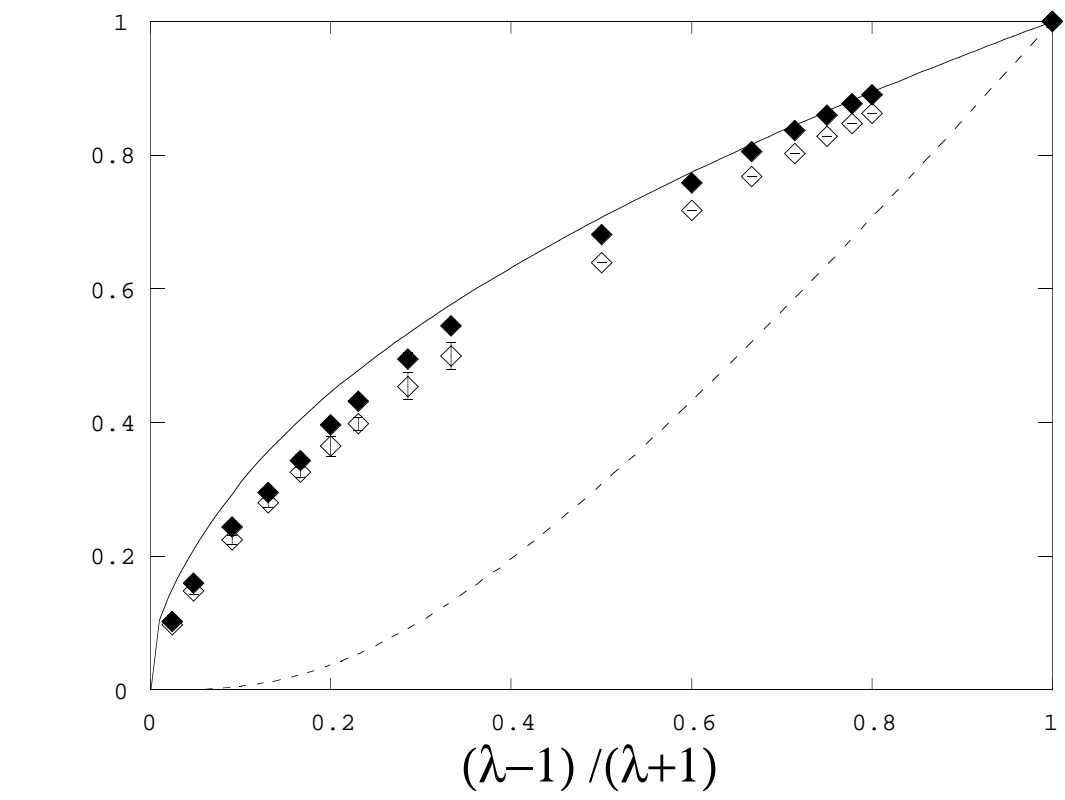
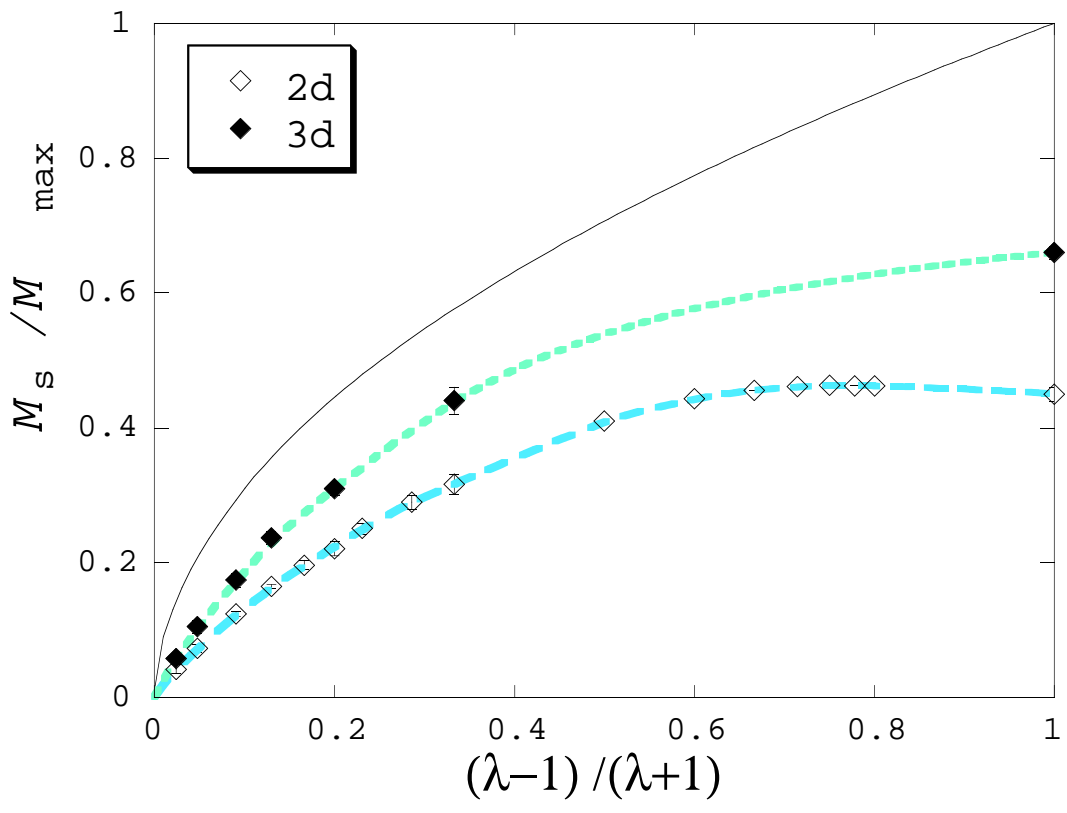
\begin{figure}
\caption{Anisotropy $\lambda$ dependence of the energy gap $\Delta_{\rm g}$ 
(solid symbols) and the critical field $H_{\rm c}$ (open symbols)
of the spin-1/2 $XXZ$ model on a square lattice.
Solid and dotted lines correspond to the classical and one-dimensional cases,
respectively.}
\label{gap_Hc}
\end{figure}

In summary, numerical results on the magnetization process 
of the spin-1/2 Ising-like $XXZ$ models have been reported. 
The spin-1/2 $XXZ$ models on square and cubic lattices show 
a first-order phase transition at some critical magnetic field 
for the anisotropic coupling constant larger than one ($\lambda>1$).  
The critical field $H_{\rm c}$ and the magnetization jump $M_{\rm s}$ 
are estimated on the basis of the Maxwell construction. 
The critical field $H_{\rm c}$ is suppressed by quantum fluctuations (Fig.\ref{hc}).
We have demonstrated that the energy gap $\Delta_{\rm g}$ is larger 
than the critical field $H_{\rm c}$ (Fig.\ref{gap_Hc}).
The anisotropy $\lambda$ dependence of the magnetization jump $M_{\rm s}$ 
is remarkably different from the classical result (Fig.\ref{ms}).
It is strongly suggested that the value of $M_{\rm s}$ in the Ising-limit 
($\lambda\rightarrow \infty$) does not coincide with 
that of the Ising model ($\lambda=\infty$) 
in finite spatial dimensions due to quantum effects (Fig.\ref{Ising}).

\narrowtext
\acknowledgments
The authors would like to thank H.Shiba for helpful discussions 
and useful comments. One of the authors (M.K.) thanks 
T. Kawarabayashi, E. Williams and D. Lidsky for reading of the manuscript.  
The exact diagonalization program is based 
on the subroutine package "TITPACK Ver.2" coded by H. Nishimori. 
Part of the calculations were performed on the Fujitsu VPP500 
and on the Intel Japan PARAGON
at Institute for Solid State Physics, Univ. of Tokyo.

\begin{figure}
\caption{Size dependence of the critical field $H_{\rm c}$ [(a)], 
the magnetization jump $M_{\rm s}$ [(b)] and the energy gap [(c)] 
$\Delta_{\rm g}$ on the two-dimensional spin-1/2 $XXZ$ model. 
Dotted lines are guides to the eye.}
\label{size}
\end{figure}

\begin{figure}
\caption{Energy gap $\Delta_{\rm g}$ 
for the two-dimensional spin-1/2 $XXZ$ model. 
Open circles and open squares denote the results obtained 
by third-order spin-wave theory cited from \protect\cite{3OSW} 
and the series expansion cited from \protect\cite{expansion}, respectively.
Solid and dotted lines correspond to the classical and one-dimensional cases,
respectively.}
\label{gap}
\end{figure}

\begin{figure}
\caption{(a) Definition of $\theta$ and $\phi$. 
(b) Schematic picture of stable spin-configurations.
(i) N{\'e}el state. (ii) spin flopping state. (iii) fully polarized state. 
(c) Magnetization curve of the classical $XXZ$ model 
with Ising-like anisotropy.}
\label{class}
\end{figure}

\begin{figure}
\caption{Magnetization curve of the two-dimensional spin-1/2 $XXZ$ model 
with Ising-like anisotropy. Open and solid symbols denote the data 
in 26-site and 36-site clusters, respectively. 
Solid (dotted) lines correspond to the phase-separated states 
determined on the basis of the Maxwell construction 
by using the data in 36-site (26-site) clusters. 
We choose $M_1-M_2=1$ or $2$ 
in the definition of the magnetic field $H$. 
The definition of the magnetic field $H$ is in the text. 
The 36-site data for $0<M/M_{\rm max}<0.6$ are obtained by quantum Monte Carlo. 
Other data are obtained by exact diagonalization.}
\label{mh}
\end{figure}

\begin{figure}
\caption{$\partial{E}/\partial{M}$ (solid symbols) 
and $(E(M)-E(M=0))/M$ (open symbols). 
The intersection point of two curves corresponds 
to the phase-separation boundary.
The data are obtained by exact diagonalization 
in a 26-site cluster ($\lambda=2$). We set $J=1$.}
\label{det}
\end{figure}

\begin{figure}
\caption{Anisotropy $\lambda$ dependence of the critical field $H_{\rm c}$ 
of the spin-1/2 $XXZ$ model. (a) Linear plot and (b) Log-log plot.
Solid and dotted lines correspond to the classical and one-dimensional cases,
respectively.
Dashed and dash-dotted lines correspond to the cases $\alpha=0.64$ and $\alpha=0.57$. 
The definition of $\alpha$ is in the text.}
\label{hc}
\end{figure}

\begin{figure}
\caption{Anisotropy $\lambda$ dependence of the magnetization jump $M_{\rm s}$ 
of the spin-1/2 $XXZ$ model. The solid line corresponds to the classical case.
Bold lines are guides to the eye. 
This figure corresponds to the phase diagram 
of the hard-core boson model with nearest-neighbor repulsion
by relating the filling $\rho$ and the interaction strength $V/t$ 
to $M/M_{\rm max}$ and $\lambda$ 
according to $\rho=(1-M/M_{\rm max})/2$ and $V/t=\lambda/2$. 
The definition of the hard-core boson model is in the text.}
\label{ms}
\end{figure}

\begin{figure}
\caption{(a) First-order perturbation energy in the Ising-limit 
on hyper-cubic lattices. We set $J=1$ as the energy unit.
The system sizes are 400 sites for 2d, 1000 sites for 3d, 1296 sites for 4d, 
1024 sites for 5d and 4096 sites for 6d. 
Dashed lines denote the tangents from $M=0$. 
The magnetization at the point of contact corresponds to $M_{\rm s}$. 
(b) Dimensionality $d$ dependence of $M_{\rm s}$ 
in the Ising-limit on hyper-cubic lattices. 
The dashed line represents $M_{\rm s}/M_{\rm max}=1/d$.}
\label{Ising}
\end{figure}


\begin{references}
\bibitem{ene_size_correction} The finite-size effects on the energy in
the phase-separated state may be much larger than those in a
homogeneous state because of macroscopic fluctuations. The
system sizes examined in this paper may be insufficient to obtain an
accurate value for the energy of the phase-separated state. Therefore
we use the Maxwell construction to determine it.
\label{ene_size_correction}
\end{references}
\end{document}